\documentclass{aastex62}

\usepackage{color}
\usepackage{amsmath}
\usepackage{multirow} 
\submitjournal{AJ}

\shorttitle{Faster-RCNN WFSATs}
\shortauthors{Peng et al.}

\begin{document}

\title{Detection and Classification of Astronomical Targets with Deep Neural Networks in Wide Field Small Aperture Telescopes}

\author[0000-0001-6623-0931]{Peng Jia}
\affil{College of Physics and Optoelectronics,  Taiyuan University of Technology, Taiyuan, 030024, China}
\affiliation{Key Laboratory of Advanced Transducers and Intelligent Control Systems, Ministry of Education and Shanxi Province, Taiyuan University of Technology, Taiyuan, 030024, China}
\email{robinmartin20@gmail.com}

\author{Qiang LIU}
\affiliation{College of Physics and Optoelectronics,  Taiyuan University of Technology, Taiyuan, 030024, China}

\author{Yongyang SUN}
\affiliation{College of Physics and Optoelectronics,  Taiyuan University of Technology, Taiyuan, 030024, China}

\begin{abstract}
Wide field small aperture telescopes are widely used for optical transient observations. Detection and classification of astronomical targets in observed images are the most important and basic step. In this paper, we propose an astronomical targets detection and classification framework based on deep neural networks. Our framework adopts the concept of the Faster R-CNN and uses a modified Resnet-50 as backbone network and a Feature Pyramid Network to extract features from images of different astronomical targets. To increase the generalization ability of our framework, we use both simulated and real observation images to train the neural network. After training, the neural network could detect and classify astronomical targets automatically. We test the performance of our framework with simulated data and find that our framework has almost the same detection ability as that of the traditional method for bright and isolated sources and our framework has 2 times better detection ability for dim targets, albeit all celestial objects detected by the traditional method can be classified correctly. We also use our framework to process real observation data and find that our framework can improve $25 \%$ detection ability than that of the traditional method when the threshold of our framework is 0.6. Rapid discovery of transient targets is quite important and we further propose to install our framework in embedded devices such as the Nvidia Jetson Xavier to achieve real-time astronomical targets detection and classification abilities. 
\end{abstract}


\section{Introduction} \label{sec:intro}
In recent years, observing astronomical targets with variable magnitude or position has become an active research area. These targets, usually known as transients, require observations with adequate spatial and temporal resolution. To satisfy these requirements, wide field small aperture telescopes (WFSATs) are commonly used \citep{burd2005pi,ping2017the, ratzloff2019building,Xu2020}. Because WFSATs have small aperture and wide field of view, we can use several of them to build an observation net in a cost effective way. In an observation net, WFSATs are normally placed in different sites or in the same site pointing to different directions. The observation net can provide acceptable sky and temporal coverage with moderate cost. Because all these WFSATs are observing the sky in sidereal tracking mode with relatively short exposure time, there are huge amount of data to be processed each night. It is a time intensive task to process these data, since some transients require immediate follow-up observations through spectroscopy or  imaging in other bands or by larger telescopes.\\

  For transient observation tasks, data processing in WFSATs normally includes three steps: extracting images of candidate astronomical targets with source extracting method (detection), classifying these candidates into different categories (classification) and cross-matching these targets with catalogs, such as the Tycho 2 Catalog for celestial objects \citep{Hog-Tycho-2}. The efficiency and the effectiveness of the first two steps usually limit the observation ability of WFSATs, because we can not report effective information about a transient candidate if we can not detect or classify it correctly.  The state of art detection method is scanning whole frame of observation images with source extracting algorithm. Based on our experience, the most robust detection algorithm is that in the SExtractor proposed by \citet{bertin1996sextractor}.\\
  
   After detection, a lot of candidate astronomical images can be extracted and classification algorithms are required to classify these candidates. In recent years, different machine learning based astronomical image classification algorithms have been developed and they have achieved higher and higher classification accuracy and recall rate \citep{romano2006supernova, tachibana2018a, gonzalez2018galaxy, Burk2019astron_rcnn, Mahabal2019ML, Duev2019Deepstreaks, Duev2019Real-bogus,Turpin2020}. For WFSATs, we have also proposed a transient classification method based on ensemble learning and neural networks \citep{Jia2019Optical}. Our method can achieve acceptable classification performance for different kinds of astronomical targets. As more and more new image classification algorithms are proposed, we could expect better classification algorithms in the future. \\

However, all astronomical targets should be detected before they can be classified. Simply increasing the performance of classification algorithm is not enough to increase the scientific output of WFSATs. Assuming even if the classification algorithm could obtain ideal classification accuracy and recall rate, targets that can not be detected by the detection algorithm would never be processed, which will limit the observation ability of WFSATs. For example, the classification method proposed in  \citet{Jia2019Optical} can achieve human-level classification ability for different kinds of celestial objects detected by the classical SExtractor based algorithms. However, some astronomical targets such as moving targets which have streak-like images, or point-like astronomical images with low signal to noise ratio (SNR) can not be detected and they will never be processed as celestial objects. If there are transients in these celestial objects, they will never be observed. Meanwhile although these targets can not be detected, our classification neural network can achieve almost the same accuracy and recall rate for these targets as that for other astronomical targets.\\ 

  This problem is probably caused by the trade-off between versatility and specificity. For general purpose source detection algorithms, we set general parameters. Then these algorithms have robust detection ability for different kinds of astronomical targets at the expense of low detection rate for a special kind of astronomical targets. For example, the detection algorithm in SExtractor uses a very elegant rule: a group of connected pixels with values exceed a predefined value will be recognized as a detection. It can give promising results for different types of sources, but for extended targets with low SNR or multiple blended targets, the detection performance will drop down. However neural network based astronomical image classification algorithms classify images according to their overall structure. They can achieve relatively higher classification accuracy and recall rate for candidates with low SNR or several candidates that are close to each other. So if we integrate the classification algorithms with the detection algorithms, we can design a framework which would increase the performance of WFSATs in transient detection tasks.\\

  There are already several state of the art detection and classification frameworks, such as the Faster R-CNN \citep{ren2017faster}, the YOLO \citep{redmon2016you} and the SSD \citep{liu2016ssd:}. For astronomical targets, \cite{gonzalez2018galaxy} propose DARKNET which uses YOLO for real-time galaxy detection and identification. \citet{Duev2019Real-bogus} proposes to use the machine learning algorithm to classify star-galaxy and separate Real/Bogus transients. \citet{Burk2019astron_rcnn} uses semantic segmentation model named Mask R-CNN \citep{he2017mask} for real-time astronomical targets detection and classification. These methods mentioned above are mainly used for general purpose sky survey telescopes and they require large amount of astronomical images which are labeled by human experts as training data. Obtaining training data is hard for general purpose survey telescopes, because it require amount of large human labeling work. For WFSATs, it would be harder, because there are a lot of WFSATs and only limited number of images obtained by them will be checked by scientists, albeit labeling these images. Besides images obtained by WFSATs have low SNR and low spatial sampling rate. A new detection and classification framework is required.\\
  
   In this paper, we propose to adopt the concept of Faster R-CNN to build an astronomical target detection and classification framework for WFSATs. Our framework uses a Feature Pyramid Network architecture to extract features from images and a modified Resnet-50 as backbone network for classification and regression. We use simulation data to test the performance of our framework and find that our framework is better in detection of extended targets or astronomical targets with low SNR. For real applications, we propose to use simulated images and real observation images to train our framework. After training, we find that our framework has better performance than the classic framework. To further increase transit detection ability of WFSATs, we further propose to install our framework in an embedded device to achieve real time observation ability.\\
    
    This paper is organized as follows. In section 2, we introduce the transient observation task and the classic detection and classification framework in WFSATs. In Section 3, We introduce our approach and we compare the performance of our framework and that of the classic framework in Section 4. In Section 5, we show how to apply our framework to real observation data obtained by WFSATs through  transfer learning, which means we use new observation data to train our neural network that are already trained with simulated images \citep{Zhuang2019}. We make our conclusions and propose our future work in Section 6.\\ 

%--------------------------------------------------------------------------------------
\section{ Astronomical Targets Detection and Classification Task and Classic Methods in WFSATs} 
In WFSATs, transient observation tasks include discovery and observation of astronomical targets with temporal-varying properties, such as astronomical targets with magnitude or position variation. Because WFSATs are low-cost, we can use several of them to build an observation net to achieve high sky and temporal coverage. To reduce total cost and maximize detection ability, our WFSATs are working in white-light mode (no filter) and there is no field de-rotator installed in them. Due to this design concept, images obtained by our WFSATs are affected by color aberration and field rotation \citep{Jia2020}, which make traditional image difference based data processing methods hard to apply \citep{Zackay2016PROPER}. So in real applications, we directly process original observation images with the following steps:\\
  1. We select effective images which are not seriously affected by clouds or malfunction of cameras with support vector machine algorithm \citep{Wangliwen}.\\
  2. We scan whole frame of selected images with source detection method to locate candidate astronomical targets. The source detection algorithm in SExtractor is commonly used and we set very low values of threshold (1.1 $\sigma$) and connected area (3 pixels) to make sure that all targets can be detected.\\
  3. All the images of candidate astronomical targets are sent to an astronomical image classification algorithm and we will classify these targets into different categories. After this step, we could obtain positions and types of all the candidate astronomical targets.\\
  4. We cross-match different candidate astronomical targets with different catalogs. Because the diameter of our telescope is small and the exposure time is short, we will only cross-match candidate astronomical targets with bright source catalogs. After this step, all the unmatched images will be identified as transient candidates and they will be reported to data center for next step observations.\\

  With the data processing framework mentioned above, WFSATs are capable to observe bright transits, such as nearby supernova, tidal disruption flares, comets, meteors, asteroids or space debris. Although the optical design of WFSATs can guarantee good image quality, normally atmospheric turbulence becomes the major aberration contribution \citep{Jia2020b}, it is very hard to find science cameras with  low noise, high speed and large number of pixels at the same time, which would lead to low spatial sampling rates in the image plane of WFSATs. Images of astronomical targets with low spatial sampling rate are very easy to be mistaken as background variation or cosmic ray.\\

  To reduce effects brought by aforementioned problems, we develop a framework which includes the detection algorithm in SExtractor and the neural network based astronomical targets classification algorithm \citep{Jia2019Optical}. It can achieve more than 94\% classification accuracy for different types of astronomical targets, but we find a problem in this framework which is caused by mismatching between the classification algorithm and the detection algorithm. The recall rate of the detection method is low for extended targets and targets with low SNR. It means many astronomical targets, which can be classified correctly, have been ruled out in the detection step. This problem is vital to WFSATs because it would reduce the detection ability of the whole observation net. Considering the machine learning based classification algorithm has already proven its good performance in classifying between different targets, we propose to develop new framework based on the classification algorithm, which can detect and classify images of different astronomical targets at the same time. In the next section, we will introduce our framework.\\

%--------------------------------------------------------------------------------------
\section{Astronomical Targets Detection and Classification Framework based on Deep Neural Networks} 
Because the spatial sampling rate in WFSATs is low (several arcsec per pixel) and the exposure time is short (around several seconds), images obtained by WFSATs are quite different from images obtained by general purpose sky survey telescopes. For ordinary stars with moderate SNR, images normally have around $5 \times 5$ pixels. For bright stars or fast moving targets, images will extend to tens of pixels. This would lead to a different classification and detection framework, comparing with the framework proposed for general purpose sky surveys.\\

\subsection{Architecture of the Astronomical Targets Detection and Classification Framework for WFSATs}
\begin{figure*}
   \centering 
   \includegraphics[width=0.9\textwidth]{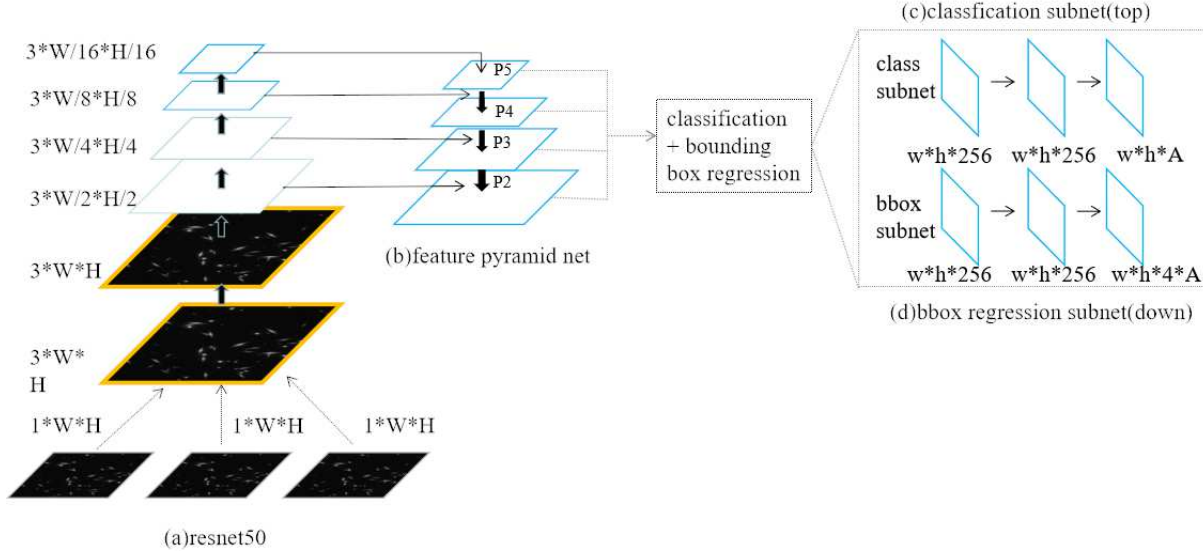}
      \caption{Architecture of the Faster R-CNN used in this paper. An input image includes three channels and image in each channel is the same astronomical image.  The input image passes into the first convolutional layer, which has three convolution layers with stride of 1 and convolution kernel of $3 \times 3$ to process each channel of the input image. Then the output of the first convolutional layer will be put into the feature pyramid network for feature extraction. In the feature pyramid network, feature maps from the P2 layer are used as features of images for each candidates. These feature maps will be used for classification through comparing between the features of different categories and they will also be used for position regression through bounding box regression. ReLU function, which uses the rectifier activation function $max(0,n)$ to evaluate outputs of previous layers, is used as activation functions for all five hidden layers. In this figure, $W$ and $H$ stands for the size of original image. $w$ and $h$ stands for the size of candidate image. The blue boxes stand for different convolutional stage and shape of these layers is shown beside these boxes. }
         \label{fig3}
   \end{figure*}

The framework proposed in this paper adopts the concept of the Faster R-CNN neural network \citep{ren2017faster}. It uses several windows with pre-defined size to scan the whole frame of an image and small images obtained by these windows will be sent into neural networks for classification and regression. The classification results will indicate different types of astronomical targets and the regression results will provide positions of these astronomical targets. The overall architecture of our framework is illustrated in Figure \ref{fig3} and it can be divided into four parts:\\

\begin{figure*}
   \centering 
   \includegraphics[width=0.9\textwidth]{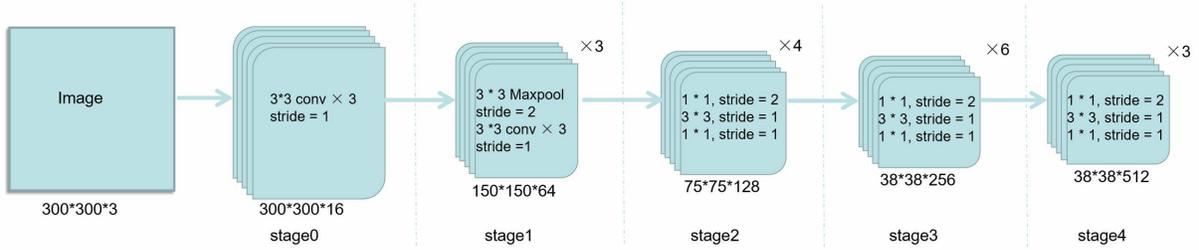}
      \caption{The structure of the Resnet50 used in this paper. It includes 5 convolutional stages shown as boxes and in each stage there are several convolutional layers. The size of convolutional kernel in these convolutional layers is shown as $n \times n$. The size of each stage is shown below each box with width$\times$length$\times$channel.}
         \label{figresnet}
   \end{figure*}

 	1. \textbf{Feature Extraction}. In this part, neural networks are used to extract features of images for subsequent Region Proposal Network layer (RPN), which will propose candidate regions that contain celestial objects for classification and position regression, and Fully Connection layer (FC). Convolutional neural networks are commonly used to extract features from images. In this paper, we choose the feature pyramid network (FPN) structure, which consists of convolutional layers, relu activation layers and pooling layers as shown in left part of Figure \ref{fig3}. Because the size of astronomical images obtained by WFSATs is small,  if we use CNN with multiple layers for feature extraction, features and spatial location will be dispersed quickly. Besides because images of candidate astronomical targets have different size as that of the perceptive filed, the classification accuracy will drop down \citep{imporved_SRN}. To solve these problems, we use a customized Resnet 50 as backbone of our neural network. Besides we use convolutional kernel with size of $3\times 3$ to replace the original convolutional kernel with size of $7\times 7$ in the Resnet-50 as shown in figure \ref{figresnet}.\\

	2. \textbf{Region Proposal}. The region proposal stands for proposal of small regions that contain celestial objects for subsequent networks. In this part, we will firstly divide the original image into small images of different size and shape with pre-defined values. These small images are called anchors. Then feature maps of anchors are extracted by the FPN as we discussed in part 1. The soft-max classifier, which is a function that normalizes input vectors into a probability distribution with probabilities proportional to the exponential of the inputs \citep{Bishop2007}, will classify these anchors into either the background or the target. Then Region Proposal Networks (RPN) will change position and size of anchors labeled as targets continuously through bounding box regression and output shifted and re-scaled anchors which have maximal possibility to be astronomical targets as proposals. Normally in this step, roughly estimation of  coordinates of these candidate targets could be obtained.\\
	
	3. \textbf{RoI Pooling / Alignment}. With region proposals and their corresponding feature maps, the RoI (Region of Interest) pooling / alignment part maps proposals with different size into the same size. Then feature maps of rescaled proposals will be sent into Full Connection(FC) Layer for classification. The RoI pooling, which transfers max values within some predefined sections to the next level, is widely used in general detection frameworks \citep{Girshick2013}, such as Fast R-CNN, Faster R-CNN or RFCN. The RoI pooling will pool the corresponding areas into feature maps of fixed size according to pre-defined selection boxes. Since the size of feature maps after pooling must be a constant, there are two quantization procedures for ROI Pooling.
    \begin{enumerate}
    \item When images are passed into the backbone network, the boundary of candidate box will be quantized to the same size of the feature maps. Information in boundary parts will be lost during this process.
    \item During the pooling process, input images will be divided into units with size of $k \times k$ pixels and information in boundary of each unit is lost.
    \end{enumerate}
    
    These two quantization processes will introduce loss of regression accuracy in candidate regions, especially for small targets. For example, an image with 0.1 pixel shift in the backbone network would introduce 1.6 pixel shift in the original image, if we quantize images to $1/16$ of the original images in the backbone network. Considering images of astronomical targets obtained by WFSATs normally only have $5\times 5$ pixels, the information loss would become a serious problem. The RoI Alignment is proposed to solve this problem \citep{he2017mask}, which includes the following modifications.
    \begin{enumerate}
    \item We will iterate over each region proposal to keep floating point boundaries accurate.
    \item The region proposal is divided into $k \times k$ units and we do not quantize boundaries of each unit.
    \item Four fixed coordinate positions are calculated in each cell and values of these four positions are calculated through Bilinear Interpolation method, which evaluates values according to the distance between known coordinate positions and their values \citep{Seiler1989}. Then we will carry out the maximum pooling operation.
    \end{enumerate}
  The first and the second modification keep the regression results accurate enough, while the third modification keeps the pooling results of the backbone network accurate. In our framework, we use the ROI alignment instead of the ROI pooling. \\
	
	4. \textbf{Classification \& Regression}. In this step, feature maps of proposals are sent into the classification neural network for classification. Meanwhile, proposals will be rescaled and shifted again through bounding box regression to achieve higher regression and classification accuracy. At last, we will output  final positions and types of these proposals. After this step, the position and type of all candidate astronomical images are obtained by our framework.\\

   \begin{figure}
   \centering
   \includegraphics[width=0.4\textwidth]{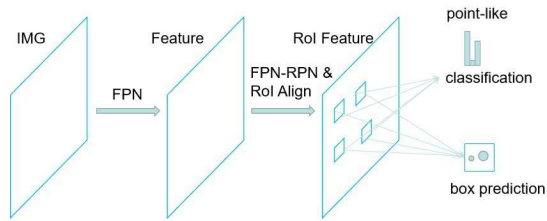}
      \caption{The flow chart of neural network used for astronomical targets detection and classification framework. The framework uses FPN to extract features from the original image. Then with RPN and ROI alignment, feature maps of candidate images are transmitted to the classification and regression neural network. Through box regression and classification, the neural network will classify and obtain position of these targets. In this figure, boxes stand for manipulation and arrows stand for data flow.}
   \label{fig1}
   \end{figure}

\subsection{Implementation of Our Astronomical Targets Detection and Classification Framework with Simulated Data}
\subsubsection{Data Preparing}
    To test the performance of our framework in detection and classification of astronomical targets, we generate mock observation data of WFSATs through Monte Carlo simulation. Because our WFSATs are working in sidereal tracking mode, there are two different astronomical targets in the observation data: point-like sources and streak-like sources. Point-like sources are normally stars, supernovae or tidal disruption flares. Streak-like sources are moving objects, such as comets, meteors, asteroids or space debris. The length of streak-like source is defined by its moving speed and exposure time and we assume the streak-like source has a moving speed of around 15-50 arcsec/s in this paper.\\
    
     We use the SkyMaker \citep{Bertin2009SkyMaker} to generate mock observation data. First of all, we assume the astronomical objects are static in very short time (around 5 ms) and generate positions and magnitudes of these targets in every instantaneous moments. Then we use SkyMaker to generate images in these moments and according to exposure time, we stack these images to generate mock observation data. In our simulation, the telescope has a field of view of 10 arcmin and pixel scale of 1 arcsec per pixel. The magnitude of sky background is 25 and the full width half magnitude of seeing disc is 1.0 arcsec. The exposure time is 1s for each frame with read out noise of 1 $e^{-}$ and dark current of 1 $e^{-}/s$. \\
    
\subsubsection{Data labeling}
\begin{figure}
   \centering
   \includegraphics[width=0.4\textwidth]{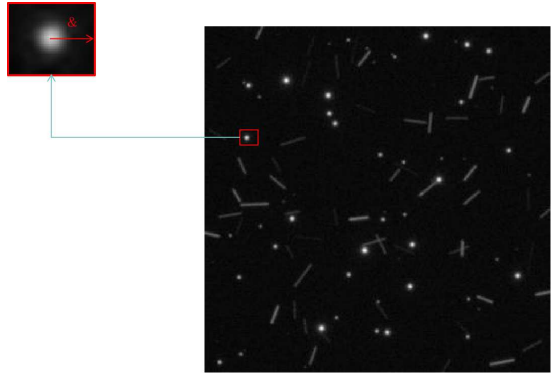}
      \caption{A ground truth box for an astronomical target. The ground truth box is created according to its center coordinate and $\sigma$. $\sigma$ is different for astronomical targets with different magnitudes.
              }
         \label{fig2}
   \end{figure}

  Labeled images which include original images and labels indicate positions and types of targets are required to train our framework. In this paper, positions of targets are labeled in original images by small rectangular box called ground truth box. For an astronomical target $i$, its ground truth box contains information stored in a vector with four dimensions: $X_{\mathrm{ih}}$, $X_{\mathrm{il}}$, $Y_{\mathrm{ih}}$ and $Y_{\mathrm{il}}$. $X_{\mathrm{ih}}$ and $X_{\mathrm{il}}$ stand for maximal and minimal values of coordinates of the target image along X direction, while $Y_{\mathrm{ih}}$ and $Y_{\mathrm{il}}$ stand for maximal and minimal values of coordinates of the target image along Y direction. Normally training data are labeled manually. However, since we use SkyMaker to generate mock observation data in this paper, we already have positions, magnitudes and types of these astronomical targets in every simulated images. According to information of these targets, we create ground truth boxes of every objects with equation \ref{equ3}. Because astronomical targets with different magnitude have different size, $\sigma$ is defined according to table \ref{tab2}.  An image of an astronomical target along with its ground truth box is shown in figure \ref{fig2}.\\
  
    \begin{eqnarray}
   \begin{aligned}
    &X_{\mathrm{ih}} &=  &X_{\mathrm{i}} + \sigma\\
    &X_{\mathrm{il}} &=  &X_{\mathrm{i}} - \sigma\\ 
    &Y_{\mathrm{ih}} &=  &Y_{\mathrm{i}} + \sigma\\ 
    &Y_{\mathrm{il}} &=  &Y_{\mathrm{i}} - \sigma 
    \end{aligned}
    \label{equ3}
    \end{eqnarray}
 
 \begin{table}
\caption{Magnitude \& Ground Truth Box Size in pixels}  
\newcommand{\tabincell}[2]{\begin{tabular}{@{}#1@{}}#2\end{tabular}}
 	 \centering
  	\begin{tabular}{c | c | c}
        \hline\hline
        Type & Magnitude & Size\\
        \hline
        \multirow{3}[6]*{point-like} & brighter than 8 & $\begin{aligned} 24 \times 24\end{aligned}$ \\
        & 9-10 & $\begin{aligned} 12 \times 12 \end{aligned}$ \\
        & 10-23 & $\begin{aligned}10 \times 10\end{aligned}$ \\
        \hline
        $\begin{aligned}
        \tabincell{c}{streak-like} \end{aligned}$ &
        $\begin{aligned}
        \tabincell{c}{19-23}\end{aligned}$ &
        $\begin{aligned}
        \tabincell{c}{
		$12 \times 12$} \end{aligned}$\\
 		 \hline 
    \end{tabular}
    \label{tab2}
\end{table} 
    
%---------------------------------------------------------------------------------------------
\subsection{Training of our framework}
  We use the following tactics to train our framework:
\begin{enumerate}
    \item Data augmentation. For real observations, data labeling is quite expensive. Before sending images into our framework, we randomly rotate them to generate several new images to increase the number of training images and generalization ability of our framework. Besides, we normalize every frame of images with equation \ref{equ5}.
    \begin{eqnarray}
    Img(i,j) = \frac{P(x_{i},y_{i}) - \overline{P(x,y)}}{P(x,y)_{max} - P(x,y)_{min}}
    \label{equ5}
    \end{eqnarray}
    where $P(x_{i},y_{i})$ is gray scale value in an image, $\overline{P(x,y)}$ is the average gray scale value of an image, $P(x,y)_{max}$ and $P(x,y)_{min}$ are the maximum and minimum gray scale values of an image respectively, $Img(i,j)$ is the image used for neural network training. 
    \item We upsample all original images to two times of their original size to increase the detection ability of our framework.
    \item We use instance normalization, which subtracts mean value and divides variance of each image for each pixel in the image, to increase converge speed during training  \citep{ulyanov2016instance}.
    \item To prevent over-fitting, we use L1 and L2 loss as regularization loss function and we set the weight of L2 loss as 0.00001 as shown in equation \ref{equL1L2},
    \begin{eqnarray}
    Loss(x,y) =\frac{1}{n}\sum_{i} \|x_i-y_i\|+0.00001*(x_i-y_i)^2.
    \label{equL1L2}
    \end{eqnarray}

    \item We choose CrossEntropy function as shown in equation \ref{equCE} as loss function for classification and smooth L1 loss function defined in \ref{equL1} for bounding box regression,\\
     \begin{eqnarray}
    L_H(p_i,y_i)=  -\sum_{k=1}^d[{{p_i}_k}\log_{{y_i}_k}+(1-{p_i}_k)\log_{(1-{y_i}_k)}] ,
    \label{equCE}
    \end{eqnarray}
where $p_{ik}$ stands for probability for celestial images of type $y_i$. The $L1_{loss}(x,y)$ stands for the smoothed L1 loss function,\\
     \begin{eqnarray}
 L1_{loss}(x,y)=\frac{1}{n}\sum_{i}z_i ,
    \label{equL1}
    \end{eqnarray}
where
\begin{equation}
 z_i=\left\{
             \begin{array}{lr}
   0.5(x_i-y_i)^2, if \|x_i-y_i\|<1\\
    \|x_i-y_i\|-0.5, otherwise
             \end{array}
\right.
\end{equation}

    \item  Anchor box is used to obtain small candidate images for further classification. The size of anchor box should be defined manually and match the size of candidate images. Because there are different types of targets for detection, the size and the aspect ratio of anchor box should be different. We use anchor box with size of [2,4,6] and aspect ratio of [0.5,1,2] in this paper. Features of images will be extracted by FPN and size of anchor box in different layers of FPN is defined as [1,2,4,8,16].
\end{enumerate}
  The whole framework is implemented with Pytorch\citep{Pytorch_inbook} in a computer with two Nvidia GTX1080Ti graphics processor units (GPU). 2000 frames of simulated data are used as training set and 500 frames of simulated data are used as test set. We initialize our neural network with random weights and train our neural network for 20-30 epochs with Adam algorithm \citep{kingma2015adam} as optimization algorithm. We set the learning rate with warming up method and the initial learning rate is set to 0.00003 \citep{zhang2019bag}.
  
\section{Performance of our framework} \label{sec:highlight}
\subsection{Performance Evaluation Method}
In this part, we will compare the performance of our framework with the classic detection and classification framework based on the SExtractor. Because the SExtractor is a source detection algorithm, we assume all the targets detected by the SExtractor can be correctly classified, which is an optimistic estimation. Furthermore, we set the detection and classification results as true positive, when the detection result is correct and the location of the detected targets is within 1.5 pixels of the ground truth position. Otherwise, we set the detection as false positive or false negative. With this definition, we can directly compare the performance of  different detection and classification frameworks. We use the mean Average Precision (mAP) to evaluate the performance of our framework and that of the classic method. Because for real observed images, it is impossible to label all astronomical targets and there are only limited targets that can be labeled by human experts, it would be more practical to use precision to evaluate the performance of our framework. The mAP is defined in equation \ref{equ7} as average accuracy of all categories,\\
   \begin{eqnarray}
   \begin{aligned}
    AP_{\mathrm{11points}} &= \frac{1}{11} \sum_{r=0}^{1.0}P_{interp}(r)\\
    mAP &= \frac{\sum_{q=1}^{Q}AP(q)}{Q}
   \end{aligned}
   \label{equ7}
   \end{eqnarray}
where $AP_\mathrm{11points}$ represent sum of the maximum precision $P_{interp}(r)$ for eleven different recall values (eg: r=0, 0.1, 0.2, 0.3, 0.4, 0.5, 0.6, 0.7, 0.8, 0.9, 1.0), when the recall value is greater than a predefined threshold. Q is the number of classes of different astronomical targets. $mAP$ is the mean value of all AP values corresponding to all categories. Because we know the position and type of all the astronomical targets, we can also use the recall rate and precision rate to evaluate the detection and classification results. The recall rate and precision rate is defined in equation \ref{equ6}.
   \begin{eqnarray}
    \begin{aligned}
    recall = \frac{TP}{TP+FN}\\
    precision = \frac{TP}{TP+FP}
    \end{aligned}
    \label{equ6}
    \end{eqnarray}
	where TP is true positives (number of correctly classified positives), TN is true negatives (number of correctly classified negatives), FP is false positives (number of incorrectly classified positives), and FN is false negatives (number of incorrectly classified negatives).\\ 

\begin{figure}
  \centering
  \includegraphics[width=0.4\textwidth]{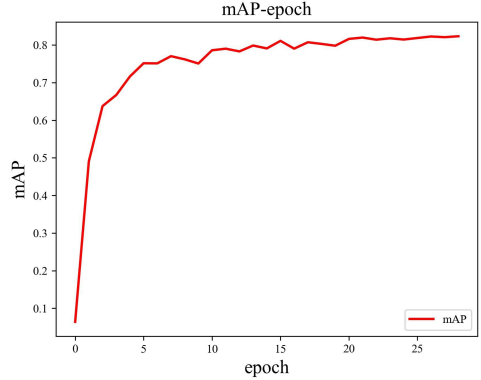}
  \caption{The Mean Average Precision curve under the condition of warming up learning rate and a test set of 500 images, We have drawn the test precision of each epoch under 30 epochs. As we can see, the value of mAP began to stabilize around 0.8 after the 30th epoch.
   }
   \label{fig7}
   \end{figure}

\begin{figure}
   \centering
   \includegraphics[width=0.4\textwidth]{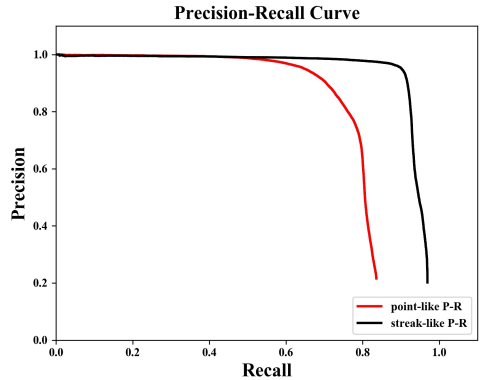}
      \caption{Precision-Recall curve for the each transient category. The red line represents the Precision-Recall curve for point-like sources, while the black curve is the Precision-Recall curve for the streak-like sources. From this figure, we can find the Faster R-CNN has different performance for different astronomical sources. The Faster R-CNN is better for detection and classification of streak-like sources.
 }
         \label{fig5}
   \end{figure}

  We will firstly show the performance of  our framework. Figure \ref{fig7} shows the curve of mAP values in the test set versus epoch number and we can find that after 30 epochs, the mAP is stable. We use 30 epochs to train our framework in real applications. After training, the precision rate versus the recall rate curve is shown in figure \ref{fig5}. We can find that the accuracy is almost the same for two different targets, while the recall rate is slightly different. Our framework is better in detecting and classifying streak-like targets.\\
	
\subsection{Comparison the performance with simulated data}	    
  In this part, we will compare the performance of our framework with that of the classic framework. We use 200 simulated images as validation set to test these two frameworks. Because we try to compare the performance of different detection and classification frameworks for targets with different SNR, we use a slightly different definition of $precision_{rate}$ and $recall_{rate}$ in this paper, as shown in equation \ref{equ8}:
	\begin{eqnarray}
   	\begin{aligned}
	precision_{rate} &=  \frac{object_{match}}{object_{total}}\frac{object_{total}}{object_{interval}}\\	
	recall_{rate} &= \frac{object_{match}}{object_{Detect}}
   	\end{aligned}
   	\label{equ8}
   	\end{eqnarray}
   	where $object_{match}$ is the number of the detected targets which are correctly detected, $object_{total}$ is the total number of object and $object_{interval}$ means the number of object  within a certain range of magnitude, and $object_{Detect}$ is the number of all the detected targets. This definition reduces problems brought by unbalanced data set. We further use $precision_{rate}$ and $recall_{rate}$ to define $f1\_score$ and $f2\_score$ to compare the performance of these two methods, as shown in equation \ref{equ9}. The $f1\_score$ is the harmonic mean  of the precision and recall, while the $f2\_score$ is weighted average of precision and recall. The $f1\_score$ is commonly used to evaluate the overall performance of classification algorithm. The $f2\_score$ is a more conservative metric, because it assumes classification accuracy is more important than recall rate. In this paper, we use both of these two metrics to evaluate the performance of these two methods.\\
   	\begin{eqnarray}
   	\begin{aligned}
	f1\_score &= \frac{2*precision*recall}{precision+recall}\\
	f2\_ score &= \frac{(1+2*2)*precision*recall}{2*2*precision+recall}
   	\end{aligned}
   	\label{equ9}
   	\end{eqnarray}

\begin{figure*}
\includegraphics[width=0.9\textwidth]{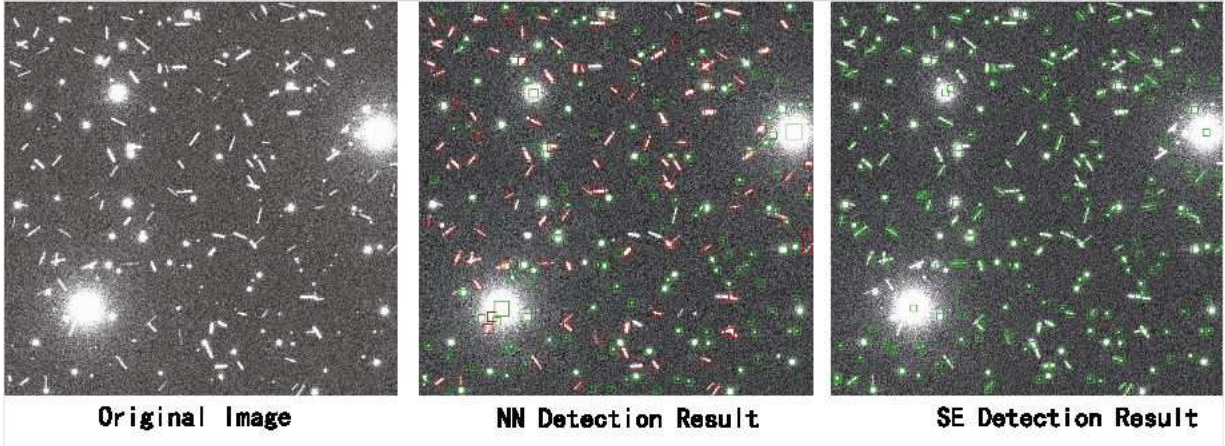}
\caption{The first image is the original image which includes both point-like sources and streak-like sources. We transform grey values in this figure to their log values for better visualization. The second image is the result obtained by our framework. Red boxes represent the streak-like astronomical images and green boxes represent point-like astronomical images. The third image is the result of detection by the classic framework. Because the classic framework can not classify targets into different types, we assume all the astronomical targets detected by the classic framework can be correctly classified. As we can see in the third image, the classic framework does not perform well for blended sources, in addition, streak-like target will be detected multiple times by the the classic framework. Therefore, when comparing the detection performance of our framework and that of the classic framework, targets that are detected by multiple times are only calculated once.}
\label{fig8}
\end{figure*}

  One frame of astronomical images in the validation set and the detection results are shown in Figure \ref{fig8}. We can find that these two methods can both detect astronomical images, but performance of these two methods is obviously different. For astronomical targets near bright target, our framework could still detect them, while the classic method is more likely to identify all these targets as one target. Meanwhile there are still several of astronomical images can not be detected by our framework. This problem requires some new methods to further increase its detection ability and we will discuss that in our next paper. \\
  
\begin{figure}
   \centering
   \includegraphics[width=0.4\textwidth]{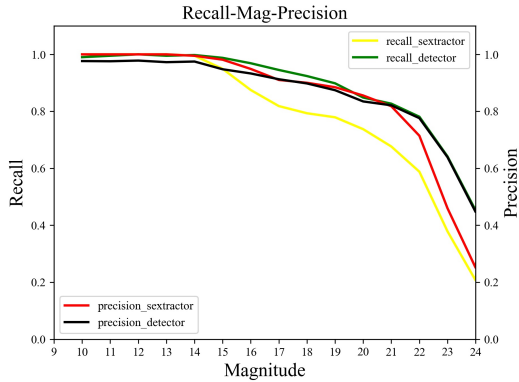}
      \caption{This figure shows how the recall rate and the precision rate will change for astronomical images with different SNR. The number of astronomical sources in each magnitude is around 2000 to 3000. From this figure, we can find the classic framework (labeled with sextractor) has slightly better performance for astronomical sources with higher SNR, while for astronomical sources with low SNR, our framework (labeled with detector) is better. }
\label{fig9}
\end{figure}

\begin{figure}
   \centering
   \includegraphics[width=0.4\textwidth]{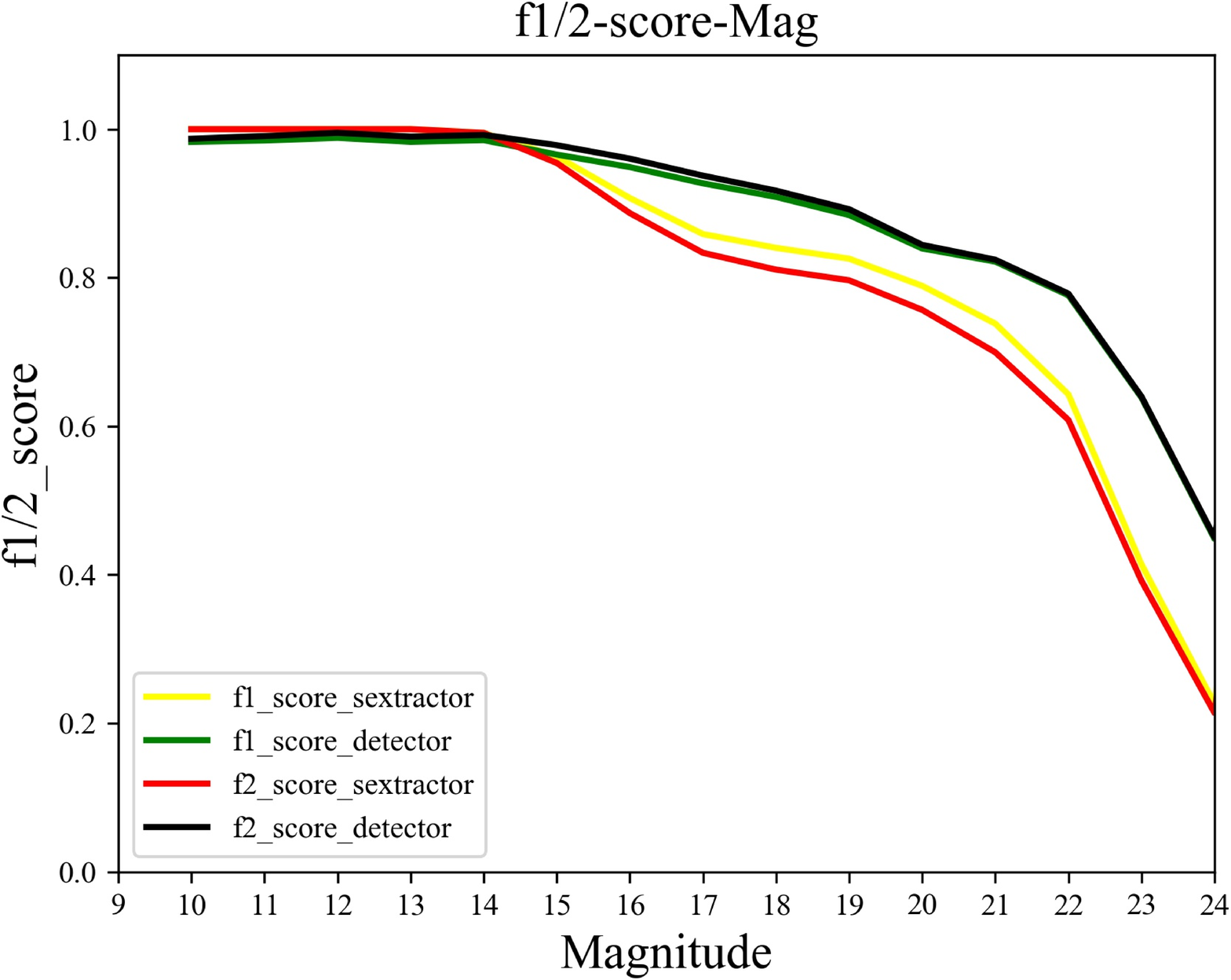}
      \caption{This figure shows the performance of these two frameworks for astronomical targets with different SNR, evaluated with $f1\_score$ and $f2\_score$. We can find that the performance of our framework (labeled with detector) is better than that of the classic framework (labeled with sextractor) for astronomical targets with low SNR and their performance is almost the same for bright sources.}
         \label{fig10}
\end{figure}

  Quantitative comparison of the performance of the classic framework and that of our framework is shown in Figure \ref{fig9}. We can find that the recall rate for bright sources are almost the same for both of these two frameworks. However for sources with low SNR, our framework have much higher recall rate. For the accuracy rate, the classic framework is slightly better than the our framework for bright sources, under the assumption that all sources detected by SExtractor can be correctly classified, which is almost impossible. When the SNR is low, the classic framework has lower classification accuracy. Overall, our framework has better performance than that of the classic framework, as shown in Figure  \ref{fig10}.\\

%-------------------------------------------------------------------------------------------
\subsection{Comparison the performance with real observation data}

In this part, we use real observation data from one of our WFSATs to test these two frameworks. This WFSAT is a refractive telescope \citep{sun2019algorithms} with parameters shown in table \ref{tab3}. To collect enough data for training and performance evaluation, we have set up a data annotation platform based on HTML and Java. Undergraduates from 6 classes are working together to annotate these images and we have collected over 600 frames of labeled images. In these images, all celestial objects are labeled and cross checked with catalogs. Then we train our framework with the following steps:\\
\begin{enumerate}
    \item Select 600 real images which contain streak-like and point-like objects.
    \item Select $75\%$ of these images as training images and the rest of them as test images.
    \item Estimate observation conditions and generate 500 frames of simulated images which also contain streak-like and point-like objects. 
    \item The simulated images and the training images are regarded as the training set, while the test images are regraded as the test set.
    \item Train our framework with the simulated images for 30 epochs as pre-train framework.
    \item Train the pre-train framework with real observation images for 10 epochs.
\end{enumerate}

\begin{table}
\caption{Details of the telescope}  
\newcommand{\tabincell}[2]{\begin{tabular}{@{}#1@{}}#2\end{tabular}}
 	 \centering
  	\begin{tabular}{c | c}
        \hline\hline
        Parameter & Value\\
        \hline
        $\begin{aligned}     
        \tabincell{c}{Aperture} \end{aligned}$ &  $\begin{aligned}  \tabincell{c}{500 mm} \end{aligned}$  \\
        \hline
        $\begin{aligned}     
        \tabincell{c}{Field of view} \end{aligned}$ &  $\begin{aligned}  \tabincell{c}{$4_{.}^{o}4 \times 4_{.}^{o}4$} \end{aligned}$  \\
        \hline
        $\begin{aligned}     
        \tabincell{c}{Size of frame(pixels)} \end{aligned}$ &  $\begin{aligned}  \tabincell{c}{$2048 \times 2048$} \end{aligned}$  \\
        \hline
        $\begin{aligned}     
        \tabincell{c}{Pixel scale} \end{aligned}$ &  $\begin{aligned}  \tabincell{c}{$7_{.}^{''}73$} \end{aligned}$  \\
        \hline
        $\begin{aligned}     
        \tabincell{c}{CCD operating mode} \end{aligned}$ &  $\begin{aligned}  \tabincell{c}{Full frame} \end{aligned}$  \\
        \hline
        $\begin{aligned}     
        \tabincell{c}{Mechanical shutter} \end{aligned}$ &  $\begin{aligned}  \tabincell{c}{None} \end{aligned}$  \\
        \hline
        $\begin{aligned}     
        \tabincell{c}{Readout channels} \end{aligned}$ &  $\begin{aligned}  \tabincell{c}{4} \end{aligned}$  \\
        \hline
    \end{tabular}
    \label{tab3}
\end{table} 

\begin{figure}
\centering
\includegraphics[width=0.4\textwidth]{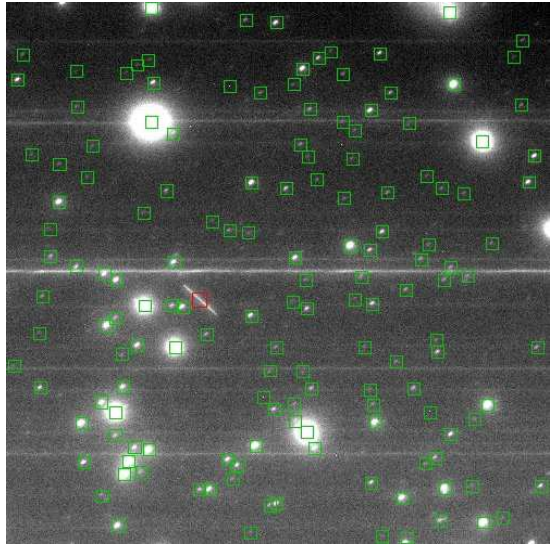}
\caption{Example of detection and classification result with real observation image. We can find our framework can correctly detect celestial objects and moving targets. Besides, hot pixels near the bright star in upper left and linear interference do not have effect to the final detection results.}
\label{fig11}
\end{figure}

After training, our framework can detect and classify astronomical targets automatically. We use the test set to test our framework and the classic framework. One frame of test images is shown in figure \ref{fig11}. We can find that our framework can detect and classify almost all astronomical targets robustly. Cosmic rays, hot pixels and linear interference do not have any effect to detection and classification results. We further compare the performance of our framework and that of the classic framework with 128 frames of validation images (images obtained by the same telescope, but in different days). Although astronomical targets in these images are checked by human experts and cross-matched by catalogs, it is still possible that there are targets not labeled. So we only show the improvement ratio of these two frameworks under different thresholds in table   \ref{tab4}. For lower threshold, our framework can give results with higher precision ratio and for higher threshold, our framework can give results with higher recall ratio. We can find that our framework can achieve around 1.032 improvement ratio, even when the threshold is as low as 0.4. If the threshold is 0.6, our framework can detect $25\%$ more targets than that detected by the classic framework.\\
\begin{table}
\caption{Improve ratio between our framework and the classic framework }  
\newcommand{\tabincell}[2]{\begin{tabular}{@{}#1@{}}#2\end{tabular}}
 	 \centering
  	\begin{tabular}{c | c}
        \hline\hline
        Threshold & Improve Ratio\\
        \hline
        $\begin{aligned}     
        \tabincell{c}{0.4} \end{aligned}$ &  $\begin{aligned}  \tabincell{c}{1.032} \end{aligned}$  \\
        \hline
        $\begin{aligned}     
        \tabincell{c}{0.5} \end{aligned}$ &  $\begin{aligned}  \tabincell{c}{1.118} \end{aligned}$  \\
        \hline
        $\begin{aligned}     
        \tabincell{c}{0.6} \end{aligned}$ &  $\begin{aligned}  \tabincell{c}{1.250} \end{aligned}$  \\
        \hline
    \end{tabular}
    \label{tab4}
\end{table}

\subsection{Applying our framework in Embedded Devices}
  To further test the performance of our framework, we use the same star catalogs to generate simulated images with different random number. Since these images are generated with the star catalog and the same level of SNR, we could expect the same detection and classification results. However we find that detection and classification results will change, even though all these images contain almost the same astronomical information. If we use all these images for object detection and classification, we can get higher accuracy rate and recall rate. This property indicates us that the best method to further increase the observation ability of WFSATs is to observe transient candidates several times, as soon as we detect them. Because there are several WFSATs in an observation net, if we can achieve real-time detection and classification ability, the observation ability of the whole observation net can be further increased.\\
  
  Based on this requirement, we propose to install our framework in an embedded device for each WFSAT to achieve real-time detection and classification ability. Since our framework uses deep neural networks, tensor cores are required to increase detection and classification speed. We install our framework in the Jetson AGX Xavier embedded device provided by Nvidia as `brain' of WFSATs. The Jetson AGX Xavier has 512-core Volta GPU and can achieve 5 TFLOPs with 16 float point accuracy. Images obtained by WFSATs are transferred into Xavier through its USB 3.1 port and it costs around 0.3 second to process an image with size of $600 \times 600$ pixels. For images with larger size, we can divide them into small images with over-lap area and process them part by part. After the detection and classification step, we can them cross-match all astronomical  targets with different catalogs in the Jetson Xavier. We will broadcast coordinates and types of candidate transients in the whole observation net. Each WFSATs can either use these information as classification and detection results from another neural network and further increase detection and classification ability through ensemble learning, or use these information as guidance to observe a particular sky area.\\ 
  
%-------------------------------------------------------------------------------------------
\section{Conclusions and Future Work} \label{sec:con}
Fast and accurate detection and classification of astronomical targets are important for transient observations. In this paper, for images obtained by WFSATs, we propose an object detection and classification framework based on deep neural networks. We compare the performance of our framework and that of the classic framework based on SExtractor with simulated and real observation data. Our framework is robust and has better performance. We further propose to install our framework in embedded devices to achieve real-time detection and classification ability, which would further increase the observation ability of the whole transient observation net. However, with simulated data, we find that our framework can not detect all the targets which indicates that there are still improvements for our framework. In our future work, we will optimize the structure of our framework to further improve the performance of our framework.

%----------------------------------------------------------------------------------------
\acknowledgments
The authors would like to thank the anonymous referee for comments and suggestions that improved the quality of this manuscript. Peng Jia would like to thank Dr. Nan Li from National Astronomical Observatories, Dr. Alastair Basden from Durham University, Professor Kaifan Ji from Yunnan Observatory, Dr. Rongyu Sun and Dr. Tinglei Zhu from Purple Mountain Observatory, Professor Qingfeng Zhang from Jinan University who provide very helpful suggestions for this paper. All the real observation images used in this paper are annotated by undergraduate students from class 1701, 1702, 1703, 1704 Opto-electronics Engineering major and class 1701 and 1702 light source and lighting major, College of Physics and Optoelectronics, Taiyuan University of Technology. This work is supported by the Joint Research Fund in Astronomy (U1631133) under cooperative agreement between the NSFC and Chinese Academy of Sciences (CAS), National Natural Science Foundation of China (NSFC)(11503018), Shanxi Province Science Foundation for Youths (201901D211081), Research Project Supported by Shanxi Scholarship Council of China, the Scientific and Technological Innovation Programs of Higher Education Institutions in Shanxi (2019L0225). Complete code can be downloaded from \url{https://doi.org/10.12149/101016}

\end{document}